\documentstyle[prc,aps,preprint,epsfig]{revtex}
\begin{document}
\draft

\title{Fusion cross sections for
superheavy nuclei in the dinuclear system concept}
\author{ G.G.Adamian$^{1,2}$, N.V.Antonenko$^{1}$,
W.Scheid$^{1}$ and V.V.Volkov$^{2}$}
\address{$^{1}$Institut f\"ur Theoretische Physik der
Justus--Liebig--Universit\"at,
D--35392 Giessen, Germany\\
$^{2}$Joint Institute for Nuclear Research, 141980 Dubna, Russia}
\date{\today}
\maketitle

\begin{abstract}
Using the dinuclear system concept we present calculations of
production cross sections for the heaviest nuclei. The obtained
results are in a good agreement with the experimental data.  The
experimentally observed rapid fall-off of the cross sections of
the cold fusion with increasing charge number $Z$ of the
compound nucleus is explained.  Optimal experimental conditions
for the synthesis of the superheavy nuclei are suggested.
\end{abstract}


\section{Introduction}
In order to reach superheavy elements and the island of
stability at $Z=114$ and $N=178-184$, two
heavy nuclei must fuse.
At the GSI (Darmstadt) the elements with $Z=$110, 111 and 112
were recently synthesized in cold fusion reactions
\cite{2}. The heaviest isotope of the element with $Z=110$ was
produced in the FLNR, JINR (Dubna) \cite{3}. The nucleus with
$Z=110$ was also produced in LBL (Berkeley) \cite{4}.
The next important step is the synthesis of the elements
with $Z=$113 and 114 by using both Pb-based \cite{5}
and actinide-based reactions \cite{6}.

The aim of investigations of the fusion mechanism is the choice
of the optimal experimental conditions for the formation
of the superheavy elements.  Producing the elements from $Z=104$
to $Z=112$ in the cold fusion reactions the experimentalists
observed the rapid fall-off of the evaporation residue cross
section (about four orders of magnitude) with increasing
$Z$ of the compound nucleus \cite{2,5,8,9}.
The measured cross section of
the production of the element with $Z=112$ is only a few pbarn.

For the cold fusion produced elements with $Z=104-112$ it was
found that the cross sections are maximal at energies below the
Bass barrier with excitation energies of the compound nuclei of
9--15 MeV \cite{2,5}. The macroscopic models in literature
\cite{10,11,12} do not reproduce the excitation function for
these reactions.  The optimal excitation energy of the compound
nucleus is much smaller than the energy predicted in models
taking a large extra-extra push energy into account \cite{12}.
In this paper we explain these effects by a fusion model
\cite{13,14} based on the dinuclear system (DNS) concept
\cite{15} and give estimations of the complete fusion
probability and optimal excitation energy for the production of
compound nuclei between $Z=104$ and $Z=114$.

In the DNS model \cite{13,14,15} the fusion process is assumed
as a transfer of nucleons from the light nucleus to the heavy
one. The DNS evolves as a diffusion process in the mass
asymmetry degree of freedom $\eta=(A_1 - A_2)/A$ to the compound
nucleus ($A_1$ and $A_2$ are the mass numbers of the nuclei and
$A=A_1 + A_2$). Evolving to the compound nucleus the DNS should
overcome the inner fusion barrier $B^*_{fus}$ in the mass
asymmetry degree of freedom. The  top of this barrier (the
Businaro-Gallone point at $\eta=\eta_{BG}$) coincides with the
maximum of the DNS potential energy as a function of $\eta$.  We
assume that complete fusion occurs after the DNS overcomes this
inner barrier.  In the DNS-concept the value of $B^*_{fus}$
represents a hindrance for complete fusion of the initial DNS
with $|\eta_i|<|\eta_{BG}|$.  Besides the motion in $\eta$ a
diffusion process in the variable of the relative distance $R$
between the DNS nuclei occurs leading to a decay of the DNS
which we denote as quasifission. For quasifission, the DNS
should overcome the potential barrier $B_{qf}$  which coincides
with the depth of the pocket in the nucleus-nucleus potential.
The energy required to overcome the fusion and quasifission
barriers is contained in the excitation energy of the DNS.

In the DNS concept the chosen potential energy surface does not
much change during the time needed for the fusion through the
Businaro-Gallone maximum in the mass asymmetry coordinate. This
maximum exists in both our and adiabatic considerations. If the
system reaches this maximum in the diffusion process, fusion
inevitably occurs much easier in asymmetrical systems
($\eta>0.75$).  Since in fusion reactions a gradual transition
between a frozen density (multinucleon transfer reactions) and
an adiabatic density (fission) happens, the complete
quantitative understanding can be reached only with a cumbersome
time-dependent calculation with a time-dependent
multidimensional potential surface. In order to estimate the
fusion cross sections in the reactions leading to the superheavy
elements in a simple manner, we use the DNS-model
\cite{13,14,15} because this model was successful in the
description of the fusion of heavy nuclei.

\section{Model}
\subsection{Evaporation residue cross section}
In accordance with the DNS-concept the evaporation
residue cross section can be written as
\begin{equation}
\sigma_{ER}(E_{\rm c.m.})=
\sum_{J=0}^{J_{max}}\sigma_{c}(E_{\rm c.m.},J)
P_{CN}(E_{\rm c.m.},J) W_{sur}(E_{\rm c.m.},J).
\label{1a_eq}
\end{equation}
The value of $J_{max}$ depends on $E_{\rm c.m.}$ and
is smaller than $J_{B_f=0}$ at which the fission barrier
in the compound nucleus vanishes.
Since the superheavy nuclei do not exist
with large angular momentum,
we can further use the following
factorization \cite{14}:
\begin{equation}
\sigma_{ER}(E_{\rm c.m.})=\sigma_{c}(E_{\rm c.m.})
P_{CN}(E_{\rm c.m.}) W_{sur}(E_{\rm c.m.}).
\label{1_eq}
\end{equation}
The capture cross section $\sigma_{c}$ defines the transition of
the colliding nuclei over the Coulomb barrier and the formation
of the DNS when the kinetic energy $E_{\rm c.m.}$ is transformed
into the excitation energy of the DNS. In the calculation of
$\sigma_{c}=\sum\limits_{J=0}^{J_{max}}\sigma_{c}(E_{\rm
c.m.},J) \approx\pi\lambdabar^2 (J_{max}+1)^2 T$ \cite{8,9}
($\lambdabar$ is the reduced de Broglie wavelength and $T$ the
transmission probability through the Coulomb barrier) for
reactions leading to superheavy nuclei, the values
$J_{max}=10\hbar$ and $T=0.5$ are used for $E_{\rm c.m.}$ near
the Coulomb barrier. The values of $\sigma_c$ obtained with this
expression are in agreement with the ones obtained in a
microscopical calculation \cite{151} based on the model
\cite{152}. The probability of complete fusion $P_{CN}$ depends
on the competition between the complete fusion and quasifission
processes after the capture stage in the DNS.  The surviving
probability $W_{sur}$ estimates the competition between fission
and neutron evaporation in the excited compound nucleus. The
competition between the complete fusion and quasifission is not
considered in the existing macroscopical models \cite{10,11,12}.

Dissipative large-amplitude collective nuclear motions,
which occur in fission, quasifission, fusion and heavy-ion
reactions, can be analyzed within the transport theory
\cite{171}.
If the initial DNS ($\eta=\eta_i$) is at the local
minimum of the potential
energy in $R$ and $\eta$, we can use a two-dimensional
Kramers-type expression (quasistationary solution of the
Fokker-Planck equation) \cite{16} for the rates of
fusion ($k=\eta$) and quasifission ($k=R$)
through the fusion ($B_\eta=B^*_{fus}$)
and quasifission ($B_R=B_{qf}$) barriers:
\begin{eqnarray}
\lambda ^{Kr}_{k}=\frac{1}{2\pi}
\frac{\omega^2_k}
{\sqrt{\omega^{B_R}_k\omega^{B_\eta}_k}}
\left(\sqrt{\left[\frac{(\Gamma/\hbar)^2}
{\omega^{B_R}_k\omega^{B_\eta}_k}\right]^2+4}
-\frac{(\Gamma/\hbar)^2}
{\omega^{B_R}_k\omega^{B_\eta}_k}\right)^{1/2}
\exp\left[-\frac{B_k}{\Theta}\right].
\label{2_eq}
\end{eqnarray}
With these rates the fusion probability can be calculated
\begin{eqnarray}
P_{CN}=\frac{\lambda_{\eta}^{Kr}}
{\lambda_R^{Kr}+\lambda_\eta^{Kr}}-
\frac{\lambda_{\eta}^{Kr}\lambda_R^{Kr}}
{\lambda_R^{Kr}+\lambda_\eta^{Kr}}
\frac{\tau_\eta-\tau_R}{\beta},
\label{3_eq}
\end{eqnarray}
where $\beta=e-1\approx 1.72$.  The first term in (\ref{3_eq})
yields the contributions of the quasistationary rates.  The
second term is related to the transient times $\tau_k$.  It was
shown in \cite{14} that we can neglect this term for $\tau_k\ll
1/\lambda_k^{Kr}$ ($k=R,\eta$) or $\tau_R\approx\tau_\eta$.
This is true for all reactions under consideration.  The
detailed discussion of Eq. (\ref{3_eq}) is given in
Ref.~\cite{14}.  The application of the Kramers-type expression
\cite{17} to relatively small barriers ($B_R/\Theta > 0.5$) was
demonstrated in \cite{18}. The local thermodynamic temperature
$\Theta$ is calculated with the expression
$\Theta=\sqrt{E^*/a}$, where $a=A/12$ MeV$^{-1}$ and  $E^*$ is
the DNS excitation energy.  In Eq.  (\ref{2_eq}), the
frequencies $\omega^{B_{k'}}_k$ ($k,k'=R,\eta$) of inverted
harmonic oscillators approximate the potential in the variables
$R$ and $\eta$ around the tops of the barriers $B_{k'}$, and
$\omega_k$ are the frequencies of the harmonic  oscillators
approximating the potential of the initial DNS. Since the
oscillator approximation of the potential energy surface is good
for the reactions considered, we neglect the nondiagonal
components of the curvature tensors in (\ref{2_eq}). The
friction coefficients are simply approximated by
$\gamma_{kk'}=\Gamma\mu_{kk'}/\hbar$ \cite{14H}.  The quantity
$\Gamma$ denotes an average double width of the single-particle
states. The calculation of the mass parameters $\mu_{RR}$ and
$\mu_{\eta\eta}$ is given in \cite{19} where it is demonstrated
that one can neglect the nondiagonal mass coefficient
$\mu_{R\eta}$ in the DNS for $|\eta|<|\eta_{BG}|$.  As shown in
\cite{14}, the friction coefficients $\gamma_{RR}$ and
$\gamma_{\eta\eta}$ obtained with $\Gamma=2$ MeV have the same
order of magnitude as the ones calculated within other
approaches.

The motion of the DNS to smaller $\eta$ leads also to
quasifission because the quasifission barrier in $R$ decreases
quickly with $\eta$ due to the increasing Coulomb repulsion.
As in Refs.~\cite{13,14,15}, we can use
the quasifission barrier $B_{qf}$ for the initial DNS
in (\ref{3_eq}) with a good accuracy.
In the considered reactions the initial DNS
is in the local minimum in $\eta$ and
$B_{qf}$ is smaller than the barrier for the motion
to smaller values of $\eta$.

\subsection{Potential energy}
The values of frequencies in (\ref{2_eq})
are easily calculated with the potential energy of the DNS
\cite{13,14}
\begin{eqnarray}
U(R, \eta, J)&= &B_1+ B_2+  V(R,\eta,J)- [B_{12}+V^{'}_{rot}(J)].
\label{7_eq}
\end{eqnarray}
Here, $B_1$, $B_2$, and $B_{12}$ are the realistic binding
energies of the fragments and the compound nucleus \cite{20},
respectively. The shell effects are included in these binding
energies. The isotopic composition of the nuclei forming the DNS
is chosen with the condition of a $N/Z$-equilibrium in the
system. The value of $U(R, \eta, J)$ is normalized to the energy
of the rotating compound nucleus by $B_{12}+V^{'}_{rot}$.  The
nucleus-nucleus potential $V(R,\eta,J)$ in (\ref{7_eq}) is
calculated as described in \cite{13,14}.  Due to the large
moments of inertia of the massive DNS considered and due to the
restricted set of angular momenta ($J\le 10\hbar$), we can
neglect the dependence of $U(R, \eta, J)$ on $J$:  $U(R, \eta,
J)\approx U(R, \eta)$, $V(R,\eta,J)\approx V(R,\eta)$.  The
calculated driving potential $U(R_m, \eta)=U(\eta)$ as a
function of $\eta$ for the reaction $^{54}$Cr+$^{208}$Pb is
presented in Fig.~\ref{1_fig} for $J=0$.  For a given $\eta$,
the value $R_m$ coincides with the position of the minimum of
the potential pocket in $V(R,\eta)$.

Deformation effects are taken into account in the calculation
of the potential energy surface \cite{14,argon}.
For the heavy nuclei in the DNS, which
are deformed in the ground state, the parameters
of deformation are taken from Ref.~\cite{21}.
The light nuclei of the DNS are assumed to be deformed only
if the energies of their $2^+$ states are
smaller than 1.5 MeV. As known from experiments on subbarrier
fusion of lighter nuclei, these states are easily populated.
For the collision energies considered,
the relative orientation of the nuclei in the DNS follows
the minimum of the
potential energy during the evolution in $\eta$.
We find that the values of $P_{CN}$ calculated with
the deformation of both nuclei in the DNS are practically the
same as the ones calculated previously in \cite{14,argon}
where a deformation only in
the heavy nucleus of the DNS was taken into account.
However, a deformation of both nuclei in the DNS yields better
agreement with the experimental excitation energies of the
compound nucleus for the $1n$ reactions.
Taking the deformation of the nuclei in the DNS as a function
of $\eta$, we choose the way for the DNS evolution on the
potential energy surface calculated in the DNS concept.

\section{Results and discussions}
In all considered cold fusion Pb-based reactions the dependence
of the potential energy of the DNS on mass asymmetry is similar
to the one presented in Fig.1. The evolution of the DNS in
$\eta$ to the compound nucleus is accompanied by overcoming the
inner fusion barrier with the height $B_{fus}^*$. Due to the
deformation of the DNS nuclei, which should be taken into
account for the 1$n$ reactions, the value of $B_{fus}^*$
decreases (the value of $P_{CN}$ increases) as compared to the
case of spherical nuclei. The values of $B_{fus}^*$ and
$B_{qf}$ are  changed from 4.8 and 4.0 MeV, respectively, for
the $^{50}$Ti+$^{208}$Pb$\to ^{258}$104 reaction till 9.0 and
1.0 MeV, respectively, for the
$^{70}$Zn+$^{208}$Pb$\to ^{278}$112 reaction.

It is important to determine the excitation energy $E_{CN}^*$
corresponding to the maximum of the excitation function in the
$1n$ fusion reaction. In order to minimize the loss because of
the fission of the excited compound nucleus, the excitation
energy should be kept as small as possible. With the value of
the inner fusion barrier, the optimal excitation energy is
calculated as $E_{CN}^*=U(R_m,\eta_i)+B_{fus}^*$
($U(R_m,\eta_i)$ is the energy of the initial DNS). Therefore,
the optimal kinetic energy is $E_{c.m.}=E_{CN}^*-Q$ which
excesses the entrance barrier by the value $\Delta
E=E_{CN}^*-B_{qf}$. Note that all considered collisions occur
above the calculated entrance barrier ($\Delta E>0$).  For
smaller and larger excitation energies, the evaporation residue
cross sections in the $1n$ fusion reaction decrease due to the
decrease of the values of $P_{CN}$ and $W_{sur}$ in
(\ref{1_eq}). The calculated optimal values of $E_{CN}^*$ (see
Fig.2) are in good agreement with the experimental data on the
$1n$ fusion reactions used in the production of the heaviest
elements with $Z=$102--112 \cite{5}.  The macroscopical models
\cite{11,12} overestimate the minimal values of $E_{CN}^*$ in
the fusion reactions. The model \cite{12} predicts $E_{CN}^*$
between 50 and 300 MeV. With the model \cite{11} the values of
$E_{CN}^*$ are estimated to be about 40--50 MeV. The use of the
Bass potential overestimates the experimental value of
$E_{CN}^*$ by 5--7 MeV \cite{5}.

The calculated values of $P_{CN}$ for the $1n$ Pb-based
reactions are presented in Fig.3.  These values are in agreement
with the ones extracted from experimental data \cite{2,5,8,9}.
The decrease of $P_{CN}$ in (\ref{1_eq}) by about four orders of
magnitude with $Z$ increasing from 104 to 112 explains the
observed rapid fall-off of the evaporation residue cross
sections. The factors $\sigma_{c}$ and $W_{sur}$ do not strongly
change with $Z$ for the cases considered.  The fusion
probability strongly decreases with decreasing mass asymmetry of
the initial DNS (increasing $Z$ for the Pb-based reactions)
because the inner fusion barrier $B_{fus}^*$ increases and the
quasifission barrier $B_{qf}$ decreases (the Coulomb repulsion
increases). For example, in the
$^{76}$Ge+$^{208}$Pb$\to ^{284}$114
reaction the estimated value of $\sigma_{ER}$ is near
the limit of present measurements.  The probability to obtain
the nucleus with $Z=116$ in the $^{82}$Se+$^{208}$Pb reaction is
smaller than this limit.  From our analysis it follows that the
fusion of symmetric combinations ($\eta_i=0$) for the synthesis
of the heaviest elements yields smaller cross sections.

Using the data in Fig.3, we can explain the smaller yield of the
nucleus with $Z=110$ in the reaction with $^{62}$Ni than the one
with $^{64}$Ni. The fusion probability in the reactions with
$^{66,68}$Zn is larger than the one with $^{70}$Zn.  However,
$W_{sur}$ in the reaction with $^{70}$Zn can be larger than
$W_{sur}$ in the reactions with other Zn isotopes because of the
smaller neutron separation energy in $^{278}$112.  It could be
that the increase of $P_{CN}$ is compensated by a decreasing
$W_{sur}$ in the reactions with the lighter isotopes.  In
addition, to obtain the same values of $\sigma_c$ for the
reactions with $^{70}$Zn and $^{68}$Zn, the excitation energy in
the reaction with $^{68}$Zn should be larger by 2 MeV than the
one with $^{70}$Zn \cite{151}. This means that
for reactions with lighter isotopes the
optimal excitation energy obtained in the static calculation
and presented in Fig.2 could be within 2 MeV smaller than
the realistic values. However, this deviation is within
the present experimental accuracy.
We note that in the reactions used for the
production of the heaviest elements all factors in (\ref{1_eq})
are equally important.

In order to calculate the evaporation residue cross sections in
the $1n$ Pb-based reactions, we use values of surviving
probabilities $W_{sur}(E^*_{CN})\approx \Gamma_n/\Gamma_f$ (the
values given refer to an angular momentum of zero) which are few
times larger than the ones estimated in \cite{8}, but smaller
than the values from the analysis of $4n$ reactions \cite{22}.
In accordance with the experimental data and shell-model
calculations \cite{23} the value of $\Gamma_n/\Gamma_f$
increases slightly for $Z=108$ because of the shell closures in
the vicinity of $N=162$. Since for larger $Z$ the neutron
separation energies and fission barriers are almost the same
\cite{23} for the compound nuclei in Table 1,
we took the same value of $\Gamma_n/\Gamma_f$ for
these nuclei. The value of $\Gamma_n/\Gamma_f$ is sensitive to
shell effects and excitation energies and has to be studied in
further details.  The calculated values of $\sigma_{ER}$
(Table 1) are in a good agreement with
the known experimental data \cite{2,5,8,9}.
One can see that at fixed $W_{sur}$ and
small change of $\sigma_c$
the value of $\sigma_{1n}$ decreases by two order of
magnitude from the nucleus with $Z=108$ to the nucleus
with $Z=113$ due to the decrease of $P_{CN}$.
Therefore, in the reaction $^{74}$Zn+$^{208}$Pb$\to^{282}$114
we expect a value of $\sigma_{1n}$ smaller than 0.1 pbarn.

In recent experiments the present limit of the heaviest
element production in the cold fusion has been reached. More
asymmetric combinations of the colliding nuclei than in the
Pb-based reactions (the initial DNS is near or behind the top of
the inner fusion barrier in mass asymmetry) can be used to
extend the production of superheavy elements. According to our
model one should take targets heavier than Pb. For the
actinide-based reactions with the projectiles like $^{48}$Ca,
$^{34,36}$S, the fusion probability is much larger than in
Pb-based reactions. This effect can compensate the increase of
the fission of the compound nucleus due to a higher excitation
energy which corresponds to the $3n-4n$ channels. Our calculated
cross section for the $4n$ reaction $^{48}$Ca+$^{244}$Pu$\to
^{292}$114 is about 1 pbarn.

\section{Summary}
In conclusion, the calculated results for the (HI,$1n$)
reactions used in the production of the heaviest elements are in
agreement with the experimental data. The calculations
for all reactions were performed with one set of
parameters and with the same assumptions.
The main factor which
prohibits the complete fusion of heavy nuclei is the
quasifission. Without regarding the quasifission, the
explanation of the experiments on the fusion of heavy nuclei is
not possible. Our model is useful in calculating the optimal
excitation energy and the combinations of the colliding nuclei.
It should be applied for the further analysis of the
experimental data.

\acknowledgments
We thank Dr. E.A.Cherepanov and Dr. A.K.Nasirov for fruitful
discussions. The author (N.V.A.) is grateful to the Alexander
von Humboldt-Stiftung for the financial support. This work was
supported in part by DFG.

\eject
\begin{table}[htbp]
\caption{Calculated [th.] and experimental [exp.] evaporation
residue cross sections for several $1n$ Pb-based reactions.
The values of $E_{\rm c.m.}$ lead to the values of $E^*_{CN}$
presented in Fig.2.
The values of $\sigma_c$, $P_{CN}$ and
$W_{sur}$ used in the calculations are explained in the text.}
\begin{tabular}{|c|c|c|c|c|c|} 
Reactions &  $P_{CN}$ & $\sigma_c$ &
$W_{sur}$ & $\sigma_{1n}$ & $\sigma_{1n}$ \\
 &   & (mb) &$\times 10^{-4}$ & [th.]& [exp.]\\\hline
$^{50}$Ti+$^{208}$Pb $\to ^{258}104$ &  3$\times 10^{-2}$ &
5.3  & 1 & 16 nb& 10 nb \\
$^{54}$Cr+$^{208}$Pb $\to ^{262}106$ &  9$\times 10^{-4}$ &
4.6 &
2 & 0.8 nb  &   0.5 nb\\
$^{58}$Fe+$^{208}$Pb $\to ^{266}108$ &  $3\times 10^{-5}$ &
4.0  &
6 & 72 pb & 70 pb \\
$^{64}$Ni+$^{208}$Pb $\to ^{272}110$ &  $1\times 10^{-5}$ &
3.4  &
6 & 20 pb & 15 pb\\
$^{70}$Zn+$^{208}$Pb $\to ^{278}112$ &  $1\times 10^{-6}$ &
3.0  &
6 & 1.8 pb &1 pb\\
$^{70}$Zn+$^{209}$Bi $\to ^{279}113$ &  $4\times 10^{-7}$ &
2.9  &
6 & 0.7 pb & \\
\end{tabular}
\end{table}


\begin{figure}
\caption{Dependence of the potential energy of the DNS on
$\eta$ for the $^{54}$Cr+$^{208}$Pb reaction ($|\eta_i|=0.59$).
The calculated results with and without deformation
of the DNS nuclei are presented by solid and dashed lines,
respectively. The result with deformation
is presented for $\eta$-values which are of interest to
calculate $B^*_{fus}$.}
\label{1_fig}
\end{figure}

\begin{figure}
\caption{ Optimal excitation energies of the compound nucleus for the $1n$
Pb-based reactions. The nuclei with even and odd $Z$ are produced
with $^{208}$Pb and $^{209}$Bi  targets, respectively.
The experimental data are shown by solid diamonds.
The projectiles are indicated.
The experimental point for $Z=112$ is shown for
$^{70}$Zn as a projectile. The calculated results are depicted by open
circles for different
projectiles. The values of $E^*_{CN}$ obtained with the Bass potential
are presented by the dashed line [4].
The solid line is drawn to guide the eye.
}
\label{2_fig}
\end{figure}

\begin{figure}
\caption{Calculated fusion probability $P_{CN}$  for different $1n$ Pb-based
reactions. The projectiles are indicated. The excitation energies
in the calculations are taken the same as in Fig. 2.}
\label{3_fig}
\end{figure}
\newpage
\epsfig{figure=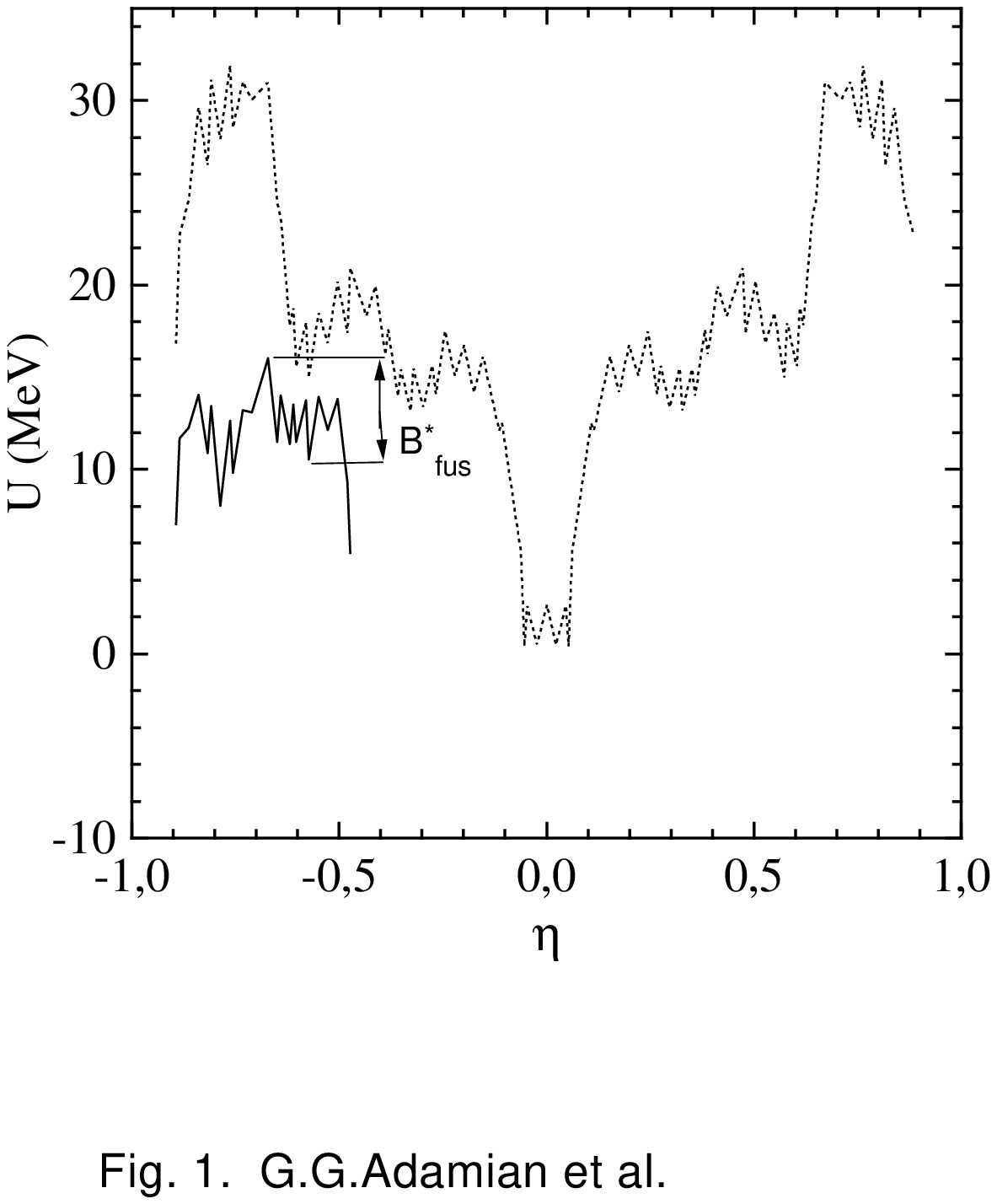,width=16cm,height=23cm}
\epsfig{figure=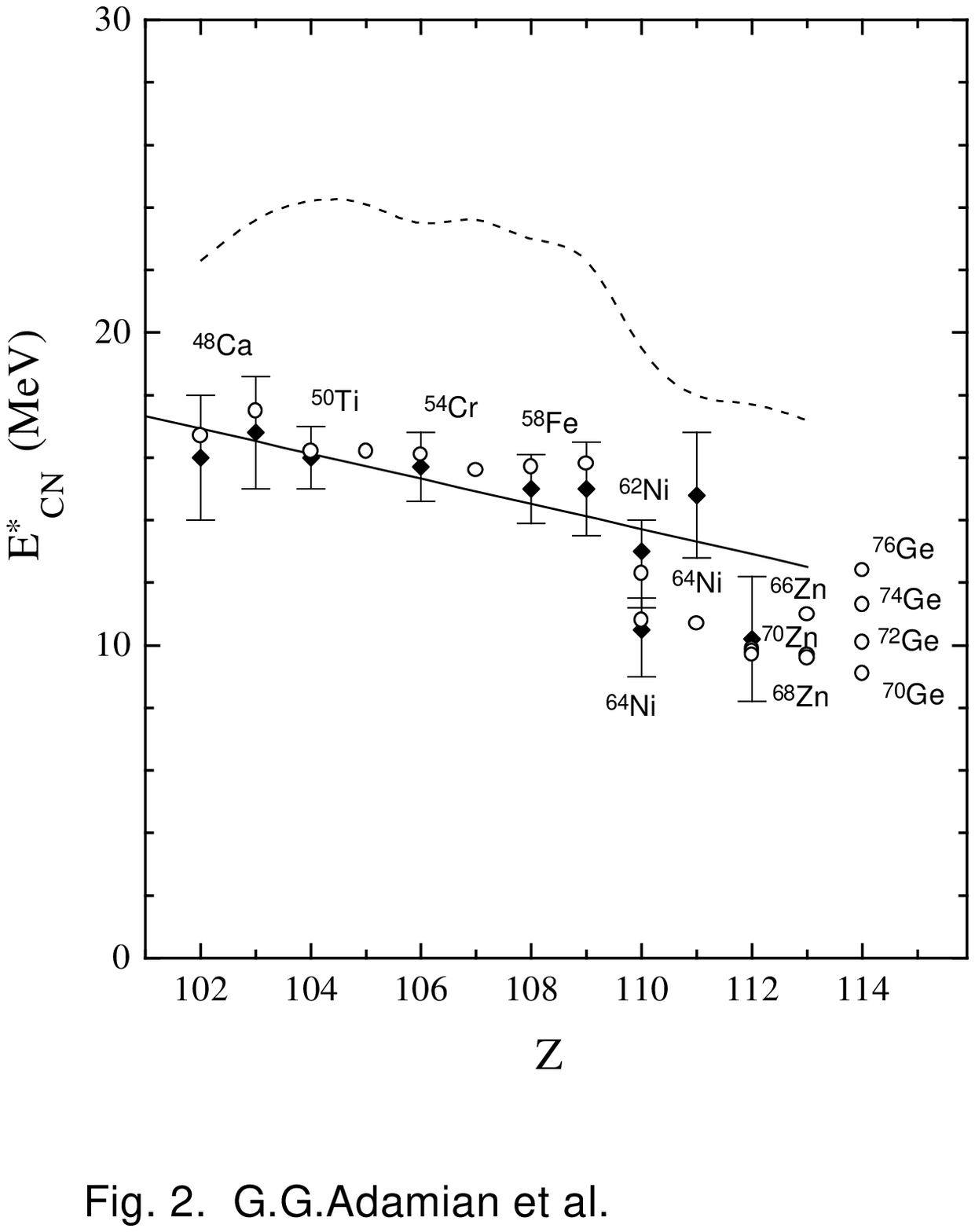,width=16cm,height=23cm}
\epsfig{figure=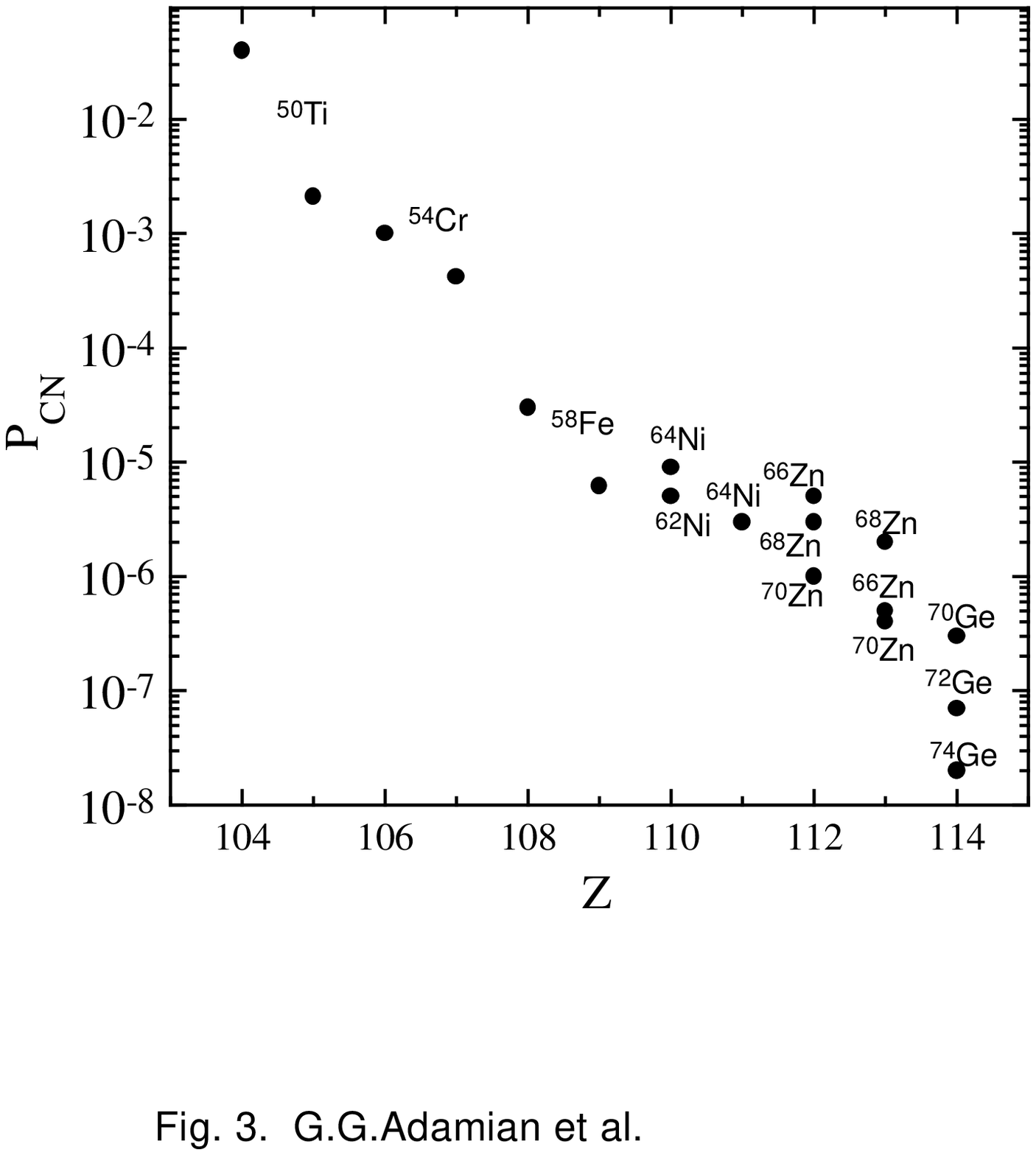,width=16cm,height=23cm}

\end{document}